\documentclass[a4paper,fleqn,usenatbib]{mnras}

\usepackage{newtxtext,newtxmath}
\usepackage[T1]{fontenc}
\usepackage{ae,aecompl}

\hypersetup{draft} % Suppressing hyperref errors preventing PDF generation as a citation is split over a page break
\usepackage[small,bf]{caption}
\usepackage{amsmath}
\usepackage{wasysym}
\usepackage{rotating}
\usepackage{multirow}
\usepackage{siunitx}
\usepackage{soul}

\newcommand{\fig}[1]{Fig.~\ref{fig:#1}}

\newcommand{\se}[1]{\S\ref{sec:#1}}

\def \ion#1#2{#1{\footnotesize{#2}}\relax}
\newcommand{\ha}{H$\alpha$}

\def \hi{\ion{H}{I}}

\title{Pipe3D Stellar and Gaseous Velocity Dispersions for CALIFA Galaxies}

\author[C. Gilhuly, S. Courteau, \& S. F. S\'anchez]{
Colleen Gilhuly,$^{1,2}$\thanks{E-mail: gilhuly@astro.utoronto.ca (CG)}
St\'ephane Courteau,$^{1}$
and Sebastian F. S\'anchez$^{3}$
\\
$^{1}$Department of Physics, Engineering Physics \& Astronomy, Queen's University, Kingston, ON K7L 3N6, Canada\\
$^{2}$Department of Astronomy \& Astrophysics, University of Toronto, Toronto, M5S 3H4, Canada\\
$^{3}$Instituto de Astronom\'ia, Universidad Nacional Aut\'onoma de M\'exico, M\'exico, D.F., M\'exico
}

\date{Accepted 2018 October 10. Received 2018 October 9; in original form 2018 June 7}

\pubyear{2018}

\begin{document}
\label{firstpage}
\pagerange{\pageref{firstpage}--\pageref{lastpage}}
\maketitle

\begin{abstract}
We present tables of velocity dispersions derived from CALIFA V1200 datacubes using 
\textsc{Pipe3D}. Four different dispersions are extracted from 
emission (ionized gas) or absorption (stellar) spectra, with two spatial apertures 
(5$''$ and 30$''$). Stellar and ionized gas dispersions are not interchangeable and 
we determine their distinguishing features. We also compare these dispersions with 
literature values and construct sample scaling relations to further assess their 
applicability.  We consider revised velocity-based scaling relations using the virial 
velocity parameter $S_{K}^2 = K V_{rot}^2 + \sigma^2$ constructed with each of 
our dispersions. Our search for the strongest linear correlation between $S_K$ and 
$i$-band absolute magnitudes favors the common $K\sim0.5$, though the range 0.3-0.8 
is statistically acceptable. The reduction of scatter in our best stellar mass-virial 
velocity relations over that of a classic luminosity-velocity relation is minimal; 
this may however reflect the dominance of massive spirals in our sample.
\end{abstract}


%%%%%%%%%%%%%%%%%%%%%%%%%%%%%%%%%%%%%%%%%%%%%%%%%%

%%%%%%%%%%%%%%%%% BODY OF PAPER %%%%%%%%%%%%%%%%%%

\section{Introduction}
\label{sec:intro}

The burgeoning field of integral field spectroscopy (IFS) is enabling spatially-resolved
 spectroscopy for unprecentedly large collections of galaxies. The Calar Alto Legacy 
Integral Field spectroscopy Area (CALIFA) survey \citep{sanchez12,walcher14,sanchez16b} 
is a forefront contributor to this movement, being the first completed IFS survey to 
target a diverse sample of galaxies in the local Universe and covering the entire Hubble 
sequence.

\textsc{Pipe3D} \citep{sanchez16, sanchez16c} is a pipeline designed to model the stellar populations 
and ionized gas of a galaxy using optical integral field unit (IFU) observations. It can 
apply to spectral data from a wide range of IFS surveys, such as MaNGA \citep{bundy15, sanchez18, abolfathi18},
SAMI \citep{bryant15}, and other IFS datasets such as the one provided by MUSE \citep{sanchezmenguiano18}
in addition to CALIFA \citep{sanchez16b, canodiaz16, bolatto17}. 
The goal of the present study is to re-examine the stellar (absorption) velocity dispersions
 obtained in the preliminary extraction steps of \textsc{Pipe3D}, prior to fitting simple 
stellar population (SSP) templates for population studies, as well as the gaseous velocity
dispersions measured during the modelling of emission lines. 

Although Pipe3D was not specifically developed for kinematic studies, the absorption and 
emission velocity dispersions are a natural by-product of the modelling process.  It is 
thus of interest to make use of these velocity dispersions to expand \textsc{Pipe3D}'s 
range of applications and assess the reliability of the extracted velocities. 

The simultaneous access to dispersions measured both in absorption and in emission for 
individual galaxies of varying morphology also enables us to examine the ``virial'' 
velocity parameter $S_K$, defined as

\begin{equation}
S_K^2 = K V_{rot}^2 + \sigma^2.
\end{equation}
 
$S_{K}$ is an alternate tracer of a galaxy's potential that takes into account contributions
from ordered rotation and random motions \citep{weiner06,kassin07}. Among others, it has 
been used to study the redshift evolution of scaling relations \citep[particularly the 
velocity-luminosity or ``Tully-Fisher'' relation of spiral galaxies][]{tully77}), since 
distant galaxies exhibit complex forms of rotational and pressure support \citep{cresci09}.
However, it is unclear whether $S_K$ applies or provides any improvements over $V_{rot}$ in 
this context \citep{miller11}. $S_K$ has also been used for the unification of local 
galaxies of diverse morphology into a common dynamical framework \citep[Aquino et al., submitted]{zaritsky08,cortese14}. 

CALIFA and other IFS surveys afford the unique ability to assess the global rotation of a 
galaxy and the local velocity dispersion from the same spectral observations for a large 
sample of galaxies. The thorough testing of various forms of $S_K$ for local galaxies, in 
order to establish best practices for future work, is thus enabled here.

The organisation of this paper is as follows. In \se{data}, we introduce our sample and data,
and compare our dispersions with one another.  Those dispersion measurements are then contrasted 
with literature analogs in \se{compare}.  Scaling relations of various forms are constructed 
in \se{scaling_relations} to assess the utility and applicability of our dispersions as 
global kinematic tracers.  We examine the virial kinematic tracer $S_{K}$ in \se{S05} in the 
context of luminosity-velocity relations in order to determine an optimal value for $K$. We 
summarize our work in  \se{summary}.

%______________________________________________________________ 

\section{Sample and data}
\label{sec:data}

Our investigation of galaxy emission and absorption velocity dispersions takes advantage
of the third CALIFA data release, employing a size-selected sample spanning
a variety of environments in the local universe (out to $z \sim 0.03$ for the main sample).
Observations were carried out at the Calar Alto Observatory with the Potsdam Multi Aperture Spectrograph (PMAS)
in the PPAK mode. The central hexagonal bundle contains 331 fibers, each 2.7$''$ in diameter,
with a total field of view of $74'' \times 64''$. An additional 36 fibers in bundles of 6 
cover the sky background. A three-pointing dithering pattern was used to cover the gaps
between the fibers and increase the final resolution of the reduced datacubes to
approximately 2.5$''$ (1 kpc at the average redshift of CALIFA galaxies).
The typical seeing was 1$''$ FWHM. 
There are two low- and medium-spectral resolution observing modes for CALIFA: V500 and 
V1200 (for grisms with 500 lines/mm and 1200 lines/mm, respectively).  The low-resolution 
V500 observations span 3745-7300\AA~ with $R \sim 850$ at 5000\AA. The medium-resolution 
V1200 observations span 3400-4750\AA~ with $R \sim 1650$ at 4500\AA. 
A total of 667 galaxies were included in the third and final CALIFA data release
\citep{sanchez12,sanchez16b,sanchez17}.

We have used \textsc{Pipe3D} \citep{sanchez16, sanchez16c} to extract our measurements of stellar
and gaseous dispersions for CALIFA galaxies, and now briefly describe the measurement procedure.
The stellar velocity dispersion, $\sigma_*$, is measured by fitting a small, representative set of stellar templates 
to the entire spectrum. The systemic velocity, $V_{sys}$, is first measured by adopting a fixed velocity dispersion for 
the convolution with the stellar templates. The instrumental width ($\sim 72$~km~s$^{-1}$) is also convolved with
the stellar templates. The systemic velocity is then held fixed while the best velocity 
dispersion is obtained by minimizing the reduced $\chi^2$.  A bounded exploration is used in both cases.
For the systemic velocity, the redshift range of CALIFA is adopted. For the velocity dispersion, we explore a range between 1-500~km~s$^{-1}$. With the 
dispersion and systemic velocity in place, the dust attenuation, $A_v$, can also be derived using a limited
set of stellar templates. Finally, knowing $V_{sys}$, $\sigma_*$, and $A_v$,  the stellar populations are then fitted 
using a Monte-Carlo approach \citep[see][for further details]{sanchez16c}. 

Emission lines are modelled with a Gaussian function and a polynomial background in the residual spectrum 
after subtracting the fitted stellar spectrum. Portions of the spectrum containing one or two emission lines 
are fitted individually rather than fitting the entire spectrum at once, to keep the form of the polynomial 
background simple. The systemic velocity and velocity dispersion of any lines fitted simultaneously
are fixed as equal. Limits to the systemic velocity of the emission lines are based on the values derived for the stellar systemic velocity, 
exploring a range of $\pm 300$~km s$^{-1}$ from that value. The allowed range of the velocity dispersion is the same as that
in the stellar case. The instrumental width is subtracted in quadrature from the emission line velocity dispersion after fitting.

\textsc{Pipe3D} yields eight different velocity dispersion measurements.
 These can be derived from either V500 or V1200 observations, fitted either in absorption (tracing stars)
or in emission (tracing ionized gas), and measured either within a 5$''$ diameter aperture 
or a 30$''$ diameter aperture. 
Due to the size-selected nature of the CALIFA sample, galaxies at higher redshift
also tend to be physically larger. Therefore, the distance covered in these apertures
is roughly constant with respect to the scale length of the galaxies.
The 5$''$ aperture covers approximately 0.2 $R_e$, and the 30$''$ aperture covers
approximately 1.2 $R_e$. 

For the gaseous velocity dispersions, the V500 lines of 
choice are [\ion{O}{III}] and \ha{} whereas the V1200 sampling favors H$\delta$. We focus 
on the V1200-based dispersion measurements, as the lower resolution V500-based dispersions 
are unreliable, especially at lower values as a result of the relatively significant 
instrumental dispersion of this observing mode. These dispersions were extracted by one of 
us (SFS) and are presented in Table~\ref{tab:sigma}. We exclude any dispersions where
the uncertainty on the measurement exceeds the measurement itself from our following analyses.
Our results change minimally if these high error entries are included.

The above dispersions are measured by fitting the sum of all spectra within the aperture.
This procedure enhances S/N, especially for the V1200 gaseous dispersions. However, if there 
is significant rotation within the aperture, the component projected along the line of sight 
will be added in quadrature with the dispersion (and instrumental dispersion) to yield a 
composite line width. In order to test the extent to which the above dispersions are 
contaminated with rotation, we have also calculated the mean of individually fitted spaxel 
dispersions (``spaxel-wise'') for the 30$''$ aperture dispersions. We assume the 5$''$ 
dispersions are minimally contaminated.

Close agreement between the spaxel-wise and standard measurements of the V1200 30$''$ aperture
stellar dispersions indicates minimal contamination from rotation. Thus, we continue to use the
standard dispersion measurement in this case. The V1200 30$''$ aperture gaseous dispersions could not
be remeasured due to insufficient S/N in the H$\delta$ line for individual spaxels.  

\begin{table*}
\begin{center}
{\small \addtolength{\tabcolsep}{5pt}
\begin{tabular}{
   c 
   S[table-number-alignment = center,
     separate-uncertainty = true,
     table-figures-uncertainty = 1,
     table-figures-decimal = 1]
   S[table-number-alignment = center,
     separate-uncertainty = true,
     table-figures-uncertainty = 1,
     table-figures-decimal = 1]
   S[table-number-alignment = center,
     separate-uncertainty = true,
     table-figures-uncertainty = 1,
     table-figures-decimal = 1]
   S[table-number-alignment = center,
     separate-uncertainty = true,
     table-figures-uncertainty = 1,
     table-figures-decimal = 1]
   S[table-number-alignment = center,
     separate-uncertainty = true,
     table-figures-uncertainty = 1,
     table-figures-decimal = 1]
   S[table-number-alignment = center,
     separate-uncertainty = true,
     table-figures-uncertainty = 1,
     table-figures-decimal = 1]
}
\hline \hline \\
\multirow{2}{*}{Name} & {\multirow{1}{*}{$5''$ $\sigma_*$}}  & {\multirow{1}{*}{$30''$ $\sigma_*$}} & {\multirow{1}{*}{$5''$ $\sigma_{H\delta}$}}
&  {\multirow{1}{*}{$30''$ $\sigma_{H\delta}$}} & {\multirow{1}{*}{$5''$ \ha{} flux}}  & {\multirow{1}{*}{$30''$ \ha{} flux}} \\ 
& {\multirow{1}{*}{(km s$^{-1}$)}} & {\multirow{1}{*}{(km s$^{-1}$)}} & {\multirow{1}{*}{(km s$^{-1}$)}} 
& {\multirow{1}{*}{(km s$^{-1}$)}} & {\multirow{1}{*}{(CGS)}}  & {\multirow{1}{*}{(CGS)}}  \\ \\ \hline \hline
\\
2MASXJ01331766+1319567 & 16.6 \pm 11.7 & 24.6 \pm 11.9 & 114.6 \pm 8.4 & 84.3 \pm 11.6 & 34.7 \pm 4.5 & 456.9 \pm 44.0 \\
ARP220 & 160.7 \pm 64.9 &  & 231.7 \pm 23.2 & 191.7 \pm 18.5 & 47.6 \pm 5.0 & 668.0 \pm 51.2 \\
CGCG163-062 & & 12.9 \pm 11.7 & 110.7 \pm 11.3 & 106.6 \pm 9.6 & 16.1 \pm 3.1 & 280.1 \pm 31.4 \\
CGCG251-041 & 142.6 \pm 44.9 & 108.1 \pm 20.7 & 171.8 \pm 16.0 & 144.2 \pm 21.6 & 231.0 \pm 14.3 & 773.6 \pm 85.4 \\
CGCG263-044 & 124.0 \pm 15.3 & 103.6 \pm 12.2 & 193.0 \pm 26.3 & 172.1 \pm 19.6 & 51.3 \pm 3.6 & 233.7 \pm 26.1 \\
CGCG429-012 & 129.2 \pm 14.3 & 124.7 \pm 12.8 &   &   & 25.3 \pm 9.6 & 203.2 \pm 82.3 \\
CGCG536-030 & 55.6 \pm 12.0 & 15.1 \pm 12.0 & 12.7 \pm 1.8 & 73.9 \pm 5.5 & 59.8 \pm 4.4 & 1122.2 \pm 56.9 \\
ESO539-G014 &  & 77.3 \pm 38.3 &  &   &   &   \\
ESO540-G003 & 83.9 \pm 12.3 & 110.2 \pm 21.4 &   & 154.0 \pm 15.7 & 24.4 \pm 12.4 & 894.9 \pm 137.9 \\
IC0159 & 206.4 \pm 188.0 & 122.9 \pm 50.9 & 56.7 \pm 3.9 & 112.6 \pm 13.5 & 1196.5 \pm 64.2 & 7087.2 \pm 607.4 \\
IC0195 & 130.9 \pm 12.4 & 96.0 \pm 12.3 &   &  & 12.8 \pm 8.4 & 82.8 \pm 77.0 \\
IC0307 & 224.3 \pm 88.0 & 165.8 \pm 24.8 & 195.8 \pm 26.7 & 257.2 \pm 15.4 & 66.0 \pm 8.6 & 501.9 \pm 86.4 \\
IC0480 & 103.8 \pm 51.6 & 98.4 \pm 69.9 & 82.6 \pm 10.8 & 130.6 \pm 17.4 & 28.0 \pm 2.4 & 589.4 \pm 47.7 \\
IC0485 & 259.0 \pm 44.2 & 163.9 \pm 17.3 & 169.9 \pm 18.1 & 249.9 \pm 12.4 & 70.4 \pm 6.5 & 407.8 \pm 46.6 \\
IC0540 & 81.1 \pm 15.4 & 69.3 \pm 12.1 & 51.7 \pm 3.7 & 111.3 \pm 10.2 & 43.3 \pm 2.9 & 228.8 \pm 35.7 \\
IC0674 & 188.6 \pm 22.1 & 179.0 \pm 13.2 & 211.8 \pm 14.7 & 227.5 \pm 28.6 & 32.3 \pm 4.1 & 335.9 \pm 51.5 \\
IC0776 & 55.9 \pm 12.9 & 8.7 \pm 11.6 & 110.3 \pm 11.2 &   & 114.4 \pm 9.1 & 1861.8 \pm 110.7 \\
IC0944 & 195.8 \pm 13.4 & 201.0 \pm 21.7 & 198.3 \pm 21.6 & 240.1 \pm 28.8 & 16.6 \pm 3.1 & 172.6 \pm 35.6 \\
IC1078 & 66.6 \pm 62.7 &  & 217.2 \pm 24.5 & 277.8 \pm 21.2 & 16.2 \pm 5.2 & 447.3 \pm 93.9 \\ \\ \hline \hline
\end{tabular} }
\caption{Table of V1200 velocity dispersions produced using \textsc{FIT3D}. V500-derived
\ha{} fluxes are also tabulated for reference. The first twenty rows of the table are shown here.
The full version of this table is available online as supplementary material, as well
as at: \url{https://www.physics.queensu.ca/Astro/people/Stephane_Courteau/gilhuly2017/index.html}.} 
\label{tab:sigma}
\end{center}
\end{table*}

\subsection{Stars versus gas}
\label{sec:gas_v_star}

% Figure 1
\begin{figure}
\begin{centering}
\includegraphics[width=0.49\textwidth]{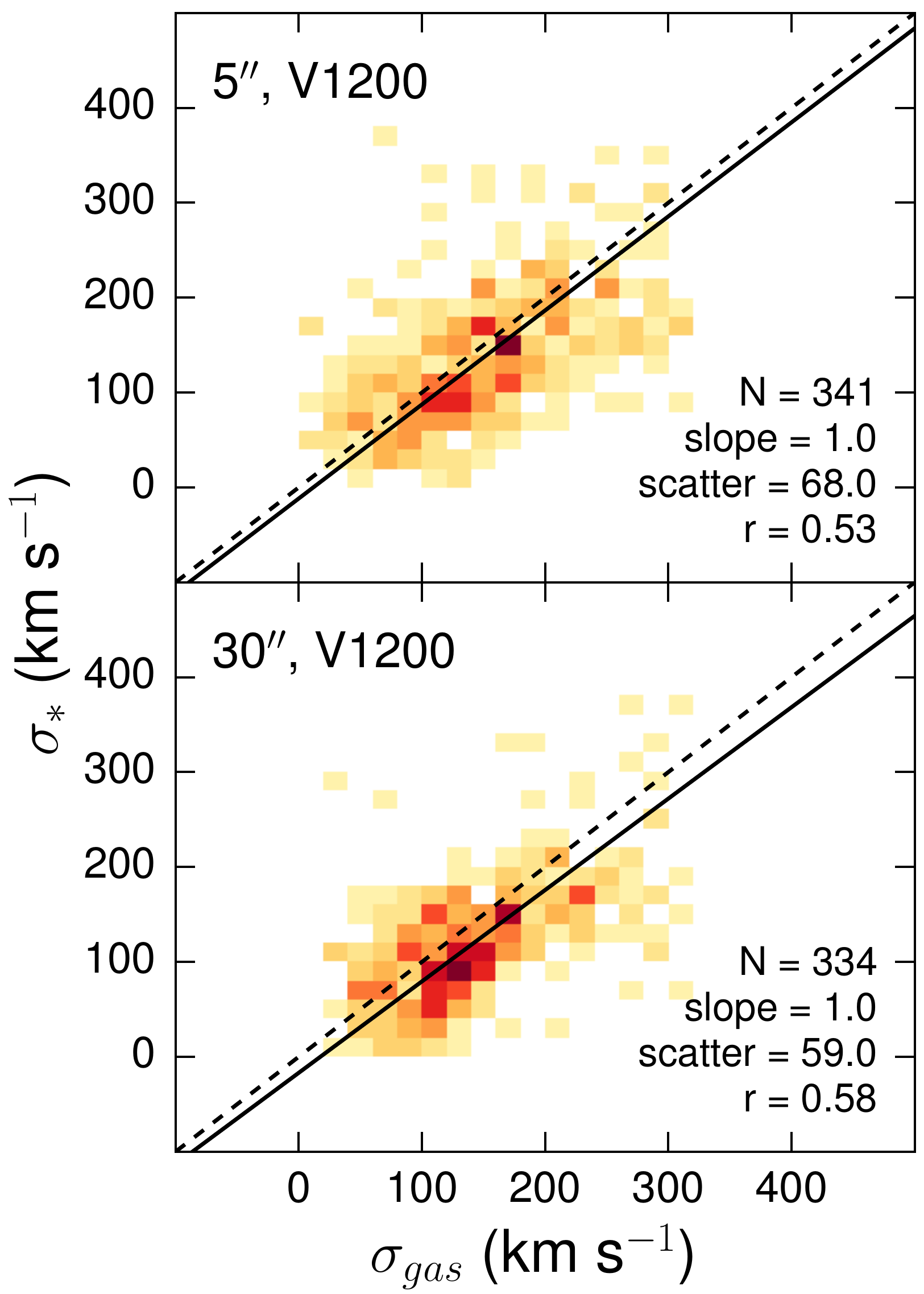}
\caption{Comparison of Pipe3D V1200 gaseous and stellar velocity dispersions.
The dashed line indicates a 1:1 relation while the solid line is the best linear fit to the data.}
\label{fig:gas_vs_star}
\end{centering}
\end{figure}

We now characterize and contrast our gaseous and stellar dispersions measured with a common aperture. 
Emission and absorption velocity dispersions are measured from tracers with typically different 
dynamical histories.The gas which gives rise to emission spectra is normally confined to the mid-plane 
of a galaxy and is dominated by rotational motions \citep[the latter may suffer from non-circular 
perturbations; see][]{spekkens07}. Gas spectra may also be broadened by non-dynamical sources such 
as hot stars, winds, turbulence, AGNs, etc. Spatially-resolved emission lines thus inform us about 
the rotational movement of its source (centroid of the spectral line) as well as the heating of its 
source (width of the line).  The spatially integrated emission spectrum, or global line width, of a 
galaxy is thus the reflection of various rotational and broadening mechanisms.

Stars may deviate from their original birth paths due to random stellar perturbations over their 
lifetime and exhibit asymmetric drift as a result.  The spatially resolved and integrated absorption 
spectra are also affected by rotational and broadening mechanisms, in a manner that differs from 
emission spectra. For instance, feedback mechanisms and winds have little effect on stellar motions.   

Given different sensitivities of stars and gas to local gravitational and thermal effects, and 
especially as those evolve over time, emission and absorption velocity dispersions are expected to 
trace different systems and thus differ from one another.  As we revisit below, the two types of 
measurements cannot be used interchangeably. 

\fig{gas_vs_star} compares the stellar and gaseous velocity dispersions, for the same aperture size,
where both dispersions are available. The correlation between these dispersions is generally weak. 
We note a tendency for galaxies with high absolute measurement error
on $\sigma$ to fall further from the best fitted line than those with low absolute measurement error.
However, the observed scatter is greater than can be explained by measurement errors alone.

In order to examine the differences between these dispersions, we consider the galaxies with the 
best and worst match amongst stellar and gaseous measurements. This is done by plotting 
$\sigma_{gas}-\sigma_*$ against various ancillary data: apparent and absolute $i$-band magnitudes 
\citep{gilhuly18}, Hubble type \citep{walcher14}, star formation rate \citep{catalan15}, mean 
stellar age at 1 $R_e$, $A$, and mean stellar metallicity at 1 $R_e$, $Z$ \citep{gonzalezdelgado15}. 
We do not find any trends in the residuals for $\sigma_{gas}-\sigma_*$ against all of the considered 
galaxy properties, suggesting a global lack of correlation between stellar and gaseous dispersions.
The latter quantities must indeed be treated as distinct entities. 

\subsection{5$''$ versus 30$''$ aperture}
\label{sec:apertures}

% Figure 2
\begin{figure}
\begin{centering}
\includegraphics[width=0.49\textwidth]{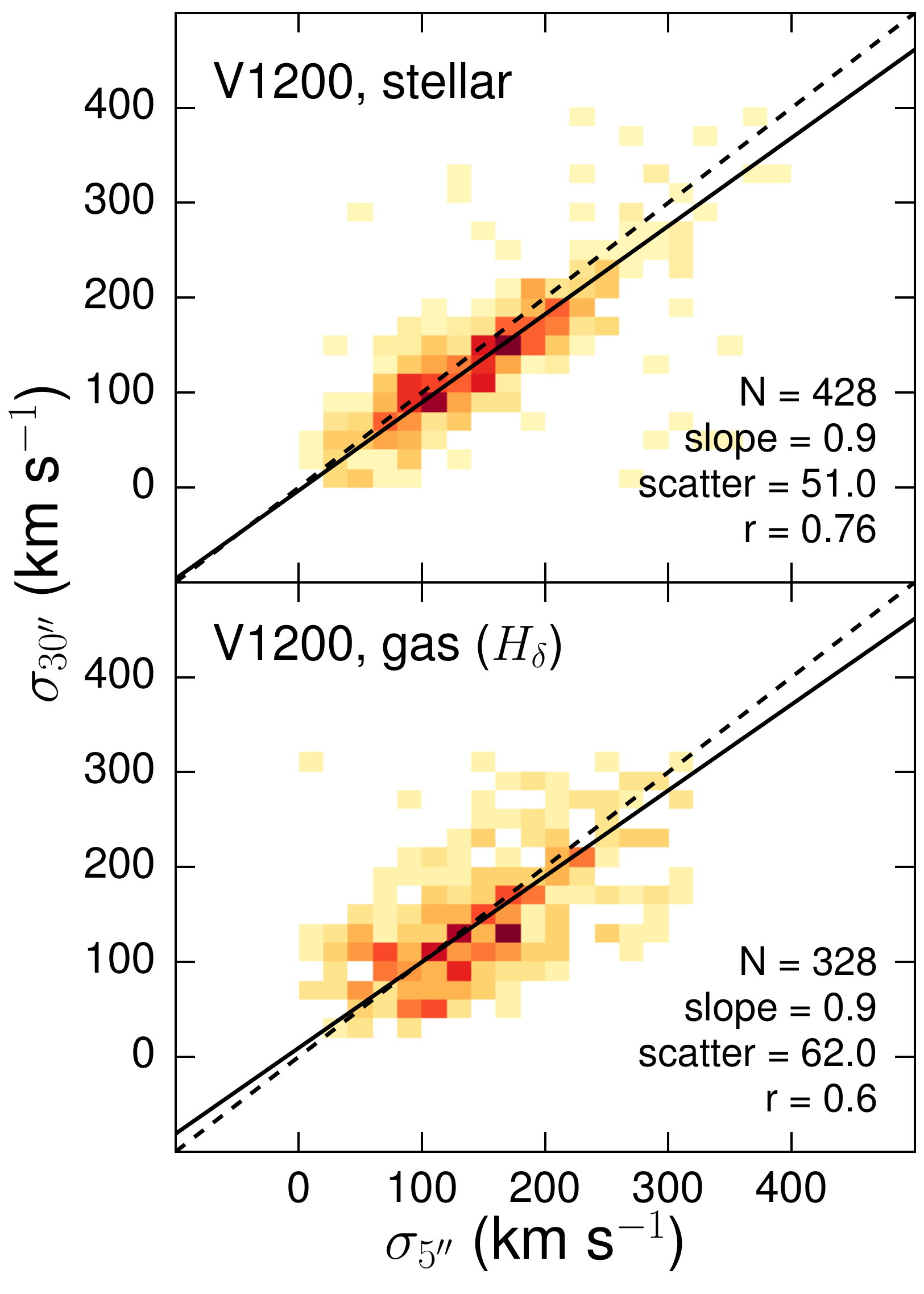}
\caption{Comparison of Pipe3D V1200 velocity dispersions measured within 5$''$ diameter and 30$''$ diameter
apertures. The dashed line indicates a 1:1 relation while the solid line is the best linear fit to the data.}
\label{fig:5_vs_30}
\end{centering}
\end{figure}

% Figure 3 (page-spanning)
\begin{figure*}
\begin{centering}
\includegraphics[width=\textwidth]{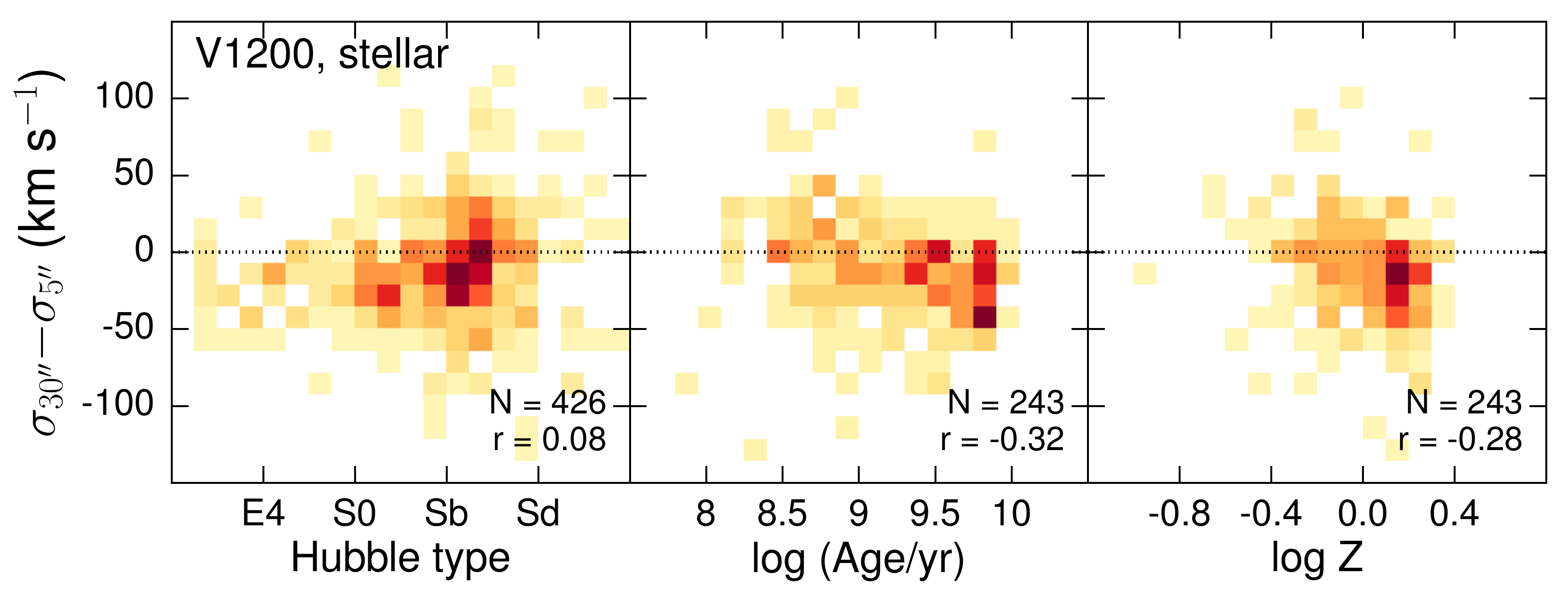}
\caption{Stellar $\sigma_{30''}-\sigma_{5''}$ versus Hubble type \citep{walcher14}, 
mean stellar age at 1 $R_e$, and mean metallicity at 1 $R_e$ \citep{gonzalezdelgado15}.}
\label{fig:5_vs_30_residuals}
\end{centering}
\end{figure*}

\fig{5_vs_30} compares dispersion measurements for 5$''$ and 30$''$ apertures. Modulo a few outliers,
the V1200 stellar dispersion shows very good agreement between our two aperture sizes for our stellar 
dispersions, with weaker agreement for the gaseous dispersions. We find a slight negative gradient in 
$\sigma_{30''}-\sigma_{5''}$ for stellar dispersions towards higher age and metallicity in \fig{5_vs_30_residuals}. 
Beyond $A < 10^9$ yrs and $Z < 1$, the scatter in $\sigma_{30''}-\sigma_{5''}$
for stellar dispersions appears to increase (signalling weaker correlation between the two dispersions).
This suggests that the stellar dispersions are less reliable for galaxies with younger stars. 
The connection  between stellar dispersions of differing apertures worsens significantly for Sc and 
later galaxies (though this may simply reflect the trends seen against age and metallicity). 

While the overall scatter in $\sigma_{30''}-\sigma_{5''}$ is greater for gaseous dispersions, we find
no connections with other galaxy properties. This indicates less consistent variation in gaseous 
dynamics from central to the mid-outer regions of CALIFA galaxies compared to stellar dynamics.  

We have demonstrated that the choice of stars versus gas and aperture size all impact the final 
measured dispersions, with especially dramatic differences between stellar and gaseous dispersions.  
In general, these dispersions cannot be used interchangeably. A more detailled investigation is 
required to determine which dispersion measure is most reliable or most informative for a given 
scientific application, as we discuss in \se{compare}.

%______________________________________________________________

\section{Comparison with existing line widths}
\label{sec:compare}

\begin{figure}
\begin{centering}
\includegraphics[width=0.49\textwidth]{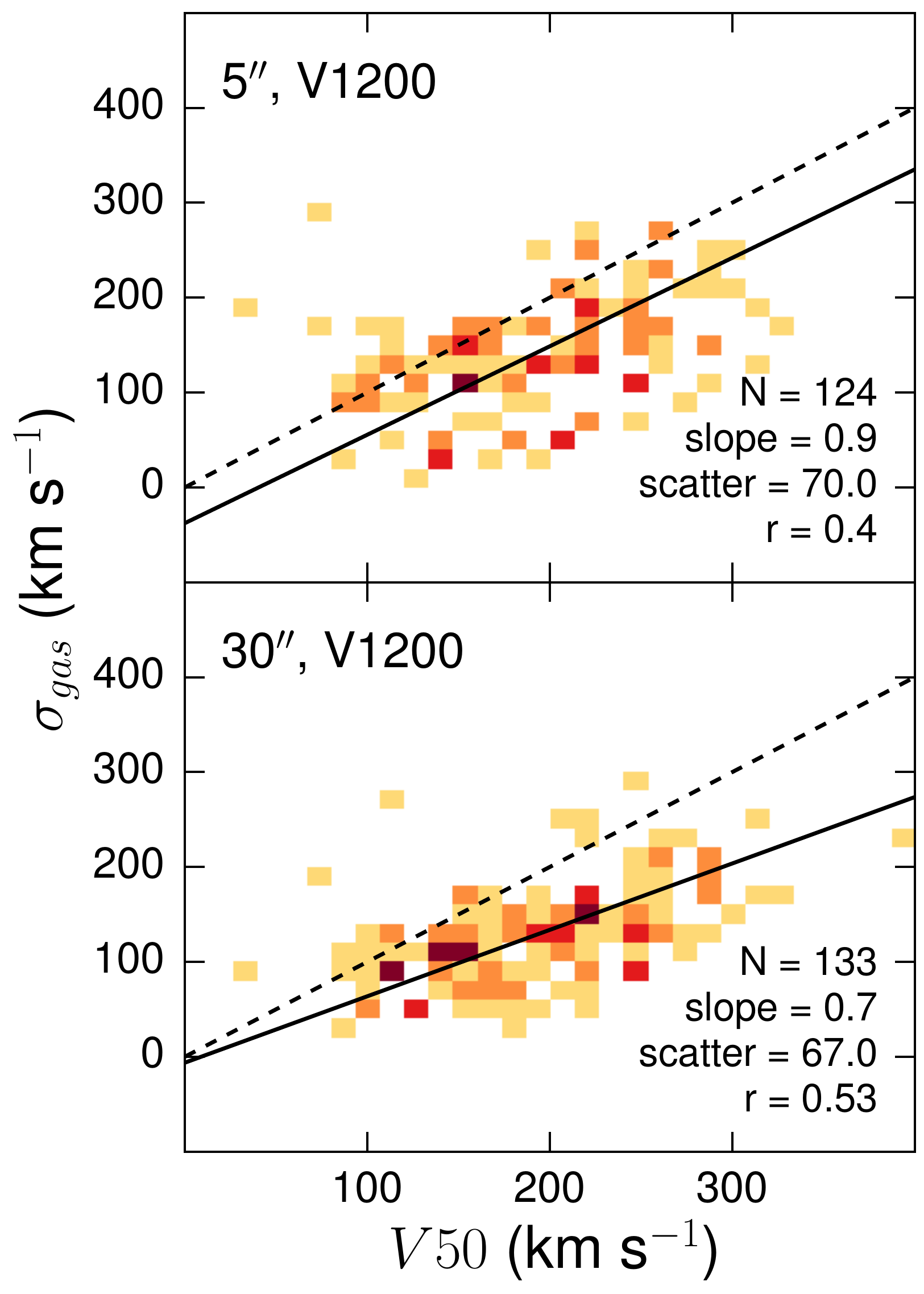}
\caption{Comparison of Pipe3D gaseous velocity dispersions with \protect\hi{} rotation velocities
derived by deprojecting and halving \protect\hi{} line widths from \protect\cite{springob05}.
The dashed line indicates a 1:1 relation  while the solid line is the best linear fit to the data.}
\label{fig:gas_vs_V50}
\end{centering}
\end{figure}

In order to validate our data and better understand their significance and applications, 
we now compare our dispersions with similar existing data products. First, we compare
our gaseous dispersions with \hi~line widths from \cite{springob05}, shown in \fig{gas_vs_V50}. 
We use W50 from their catalog, defined as the line width measured at 50\% of the peak flux. 
The peak flux and half-peak flux are measured separately for each half of the \hi~line profile, 
to better handle asymmetrical profiles.
Dispersion, instrumental, and cosmological contributions to W50 have already been removed, 
leaving a (projected) measurement of the ordered rotation of a galaxy. We deproject and halve 
the corrected line widths to obtain an estimate of rotational velocity, V50.

\begin{figure}
\begin{centering}
\includegraphics[width=0.49\textwidth]{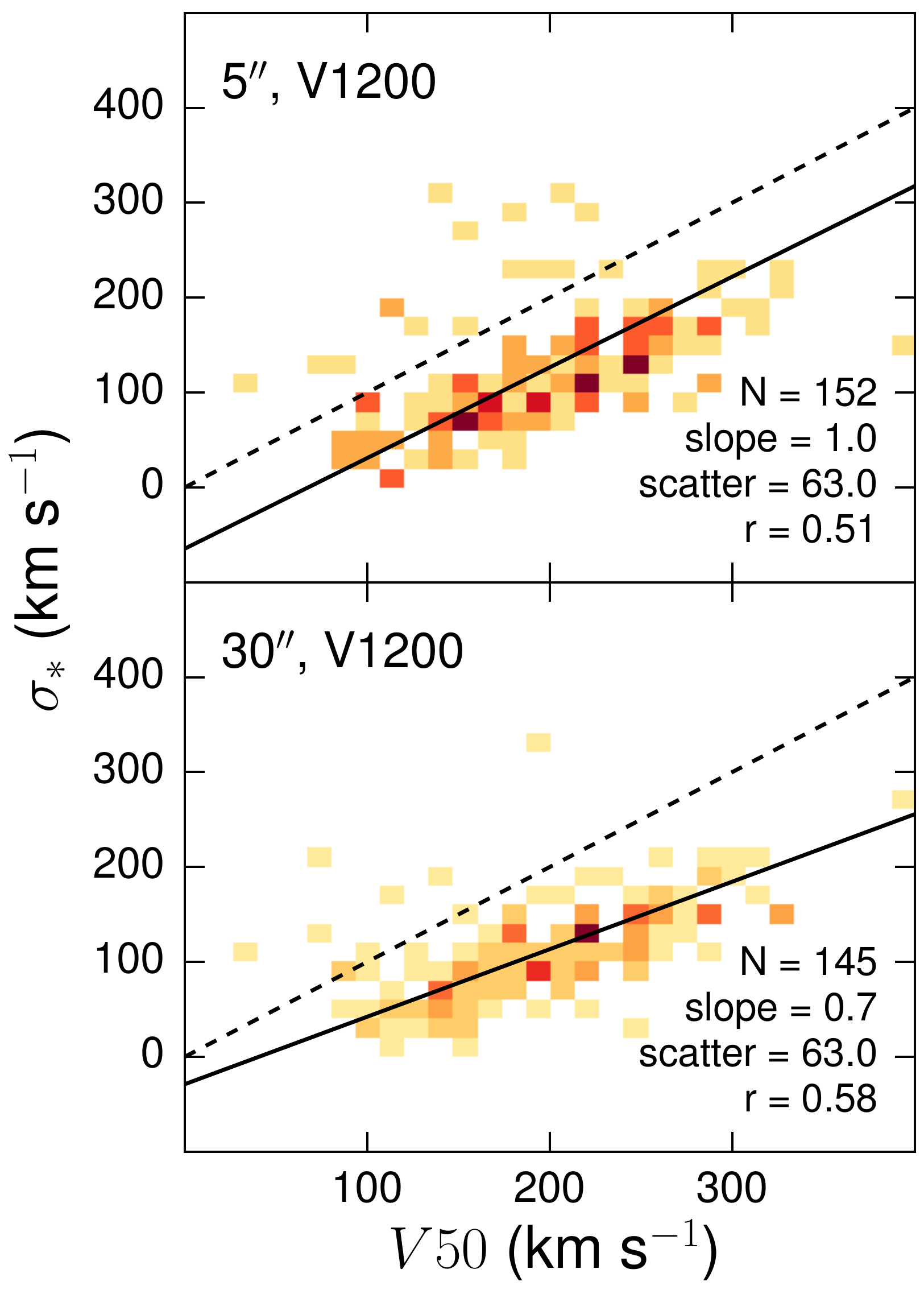}
\caption{Comparison of Pipe3D stellar velocity dispersions with \protect\hi{} rotation velocities
derived by deprojecting and halving \protect\hi{} line widths from \protect\cite{springob05}.
The dashed line indicates a 1:1 relation  while the solid line is the best linear fit to the data..}
\label{fig:star_vs_V50}
\end{centering}
\end{figure}

We find a modest correlation between our dispersions ($\sigma$) and V50, as shown in 
Figs.~\ref{fig:gas_vs_V50}~and~\ref{fig:star_vs_V50}. The dispersions are expectedly smaller 
than V50, by $\sim 50-100$ km~s$^{-1}$. Since the V50 estimates are measured at radii that are 
larger than the dispersion apertures, the interpretation of this correlation and whether it 
suffers any contamination from rotation is challenging.  Besides the fact that rotational 
motions dominate over dispersion in disc galaxies, the lower values of $\sigma$ relative to V50 
may also be due to the smaller apertures over which dispersions are measured. At the very least, 
the linear correspondence between $\sigma$ and V50 indicates that both are good tracers of the 
gravitational potential in these galaxies. The somewhat stronger correlation of V50 with 
$\sigma_*$ (relative to $\sigma_{gas}$) may reflect the weaker S/N of the H$\delta$ lines 
compared to the stellar absorption lines in the V1200 datacubes.

% Figure 4

% Figure 5

\begin{figure}
\begin{centering}
\includegraphics[width=0.49\textwidth]{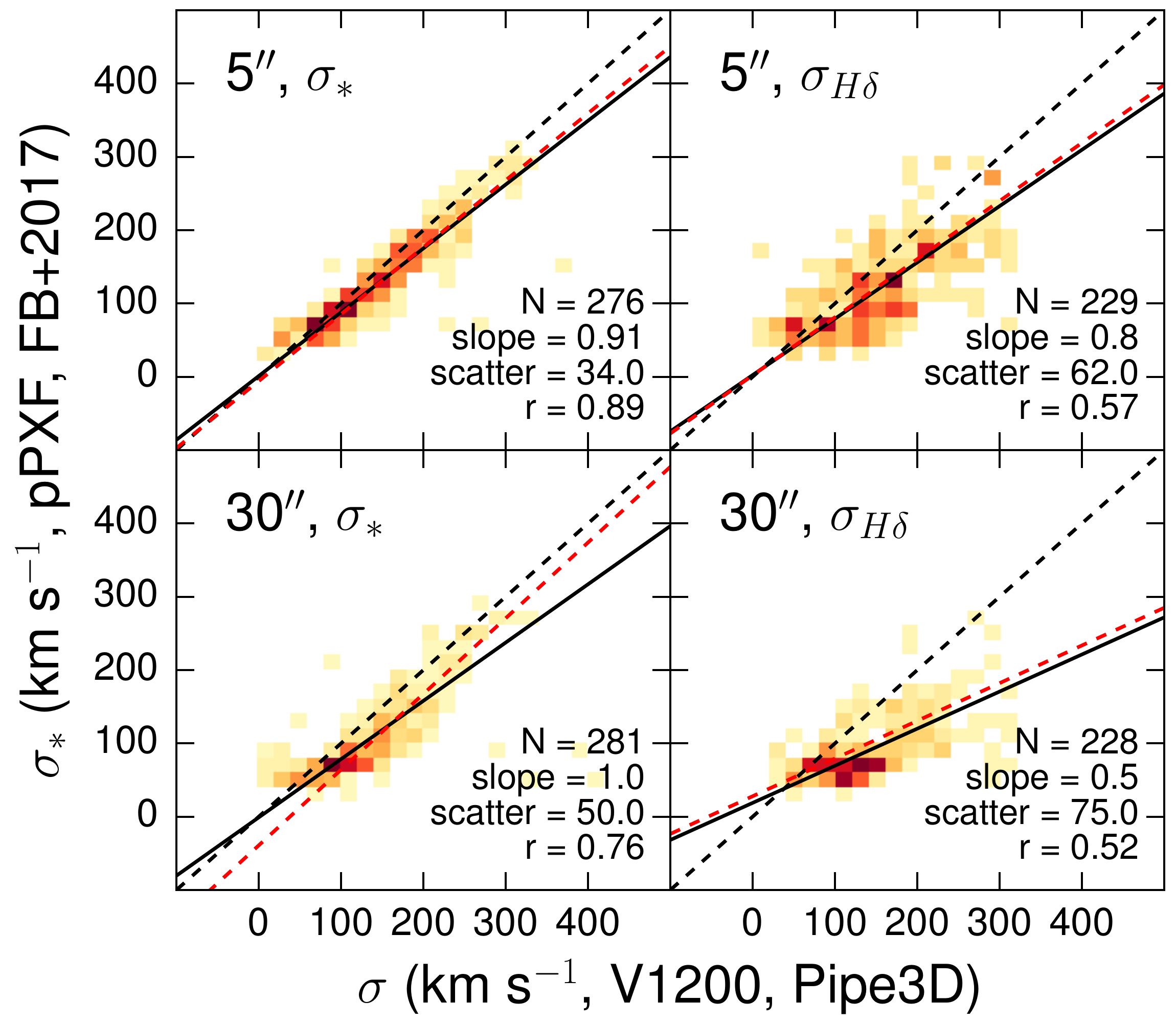}
\caption{Comparison of Pipe3D velocity dispersions with the average velocity dispersion within
comparable apertures from the maps generated by pPXF \protect\citep{falcon17}. 
The dashed line indicates a 1:1 relation while the solid line is the best linear fit to the all 
of the data. Due to the imposed minimum of 40 km/s in the pPXF maps, we also perform a secondary
fit only on galaxies where both Pipe3D and pPXF measure $\sigma > \sigma_{inst}$ ($\sim 72$ km/s).
This fit is shown in dashed red, and its best fitting parameters are displayed in each panel. We note
that the secondary fit is only significantly different from the first fit in the lower right panel,
for the comparison with our 30$''$ aperture stellar dispersions.}
\label{fig:pPXF_vs_pipe3D_v1200}
\end{centering}
\end{figure}

\fig{pPXF_vs_pipe3D_v1200} compares our V1200 dispersions with the (stellar) dispersion maps of 
\cite{falcon17} based on CALIFA V1200 datacubes. We derive approximate aperture dispersions from 
the maps by taking the average dispersion of all cells with centres within 2.5$''$ or 15$''$ of 
the centre of the galaxy. The averages are taken from Voronoi bins and are therefore flux-weighted. 
Strong agreement is found with our V1200 5$''$ stellar dispersions, with a reasonable match for our 
30$''$ stellar dispersions. Our approach to extracting an aperture dispersion and the necessarily 
larger size of the Voronoi bins relative to our apertures at greater separation from the galaxy 
centre may explain the somewhat poorer correlation between our 30$''$ stellar dispersions relative 
to the 5$''$ stellar dispersions. Establishing this favourable comparison reflects the quality of 
our dispersions. While a direct correspondence between stellar and gaseous dispersions 
(\se{gas_v_star}) is not expected, especially if radiative effects are strong, a reasonable 
correlation between our gaseous dispersions and the stellar dispersions of \cite{falcon17} is still 
found, particularly for the smaller 5$''$  aperture.

% Figure 6

%______________________________________________________________

\section{Scaling relations}
\label{sec:scaling_relations}

We pursue our validation and comparison of our velocity dispersions by examining a selection of 
dynamical scaling relations. The tight correlation of independent galaxy properties with our 
velocity dispersions may further enhance our understanding of the physical processes governing 
galaxy formation and evolution. We make use of the extensive CALIFA photometry catalog of 
\cite{gilhuly18} for total $i$-band magnitudes, effective surface brightnesses and radii, colours, 
stellar masses, and colours in order to construct various scaling relations and to verify for completeness.

\subsection{Fundamental Plane}
\label{sec:FP}

\begin{figure}
\begin{centering}
\includegraphics[width=0.49\textwidth]{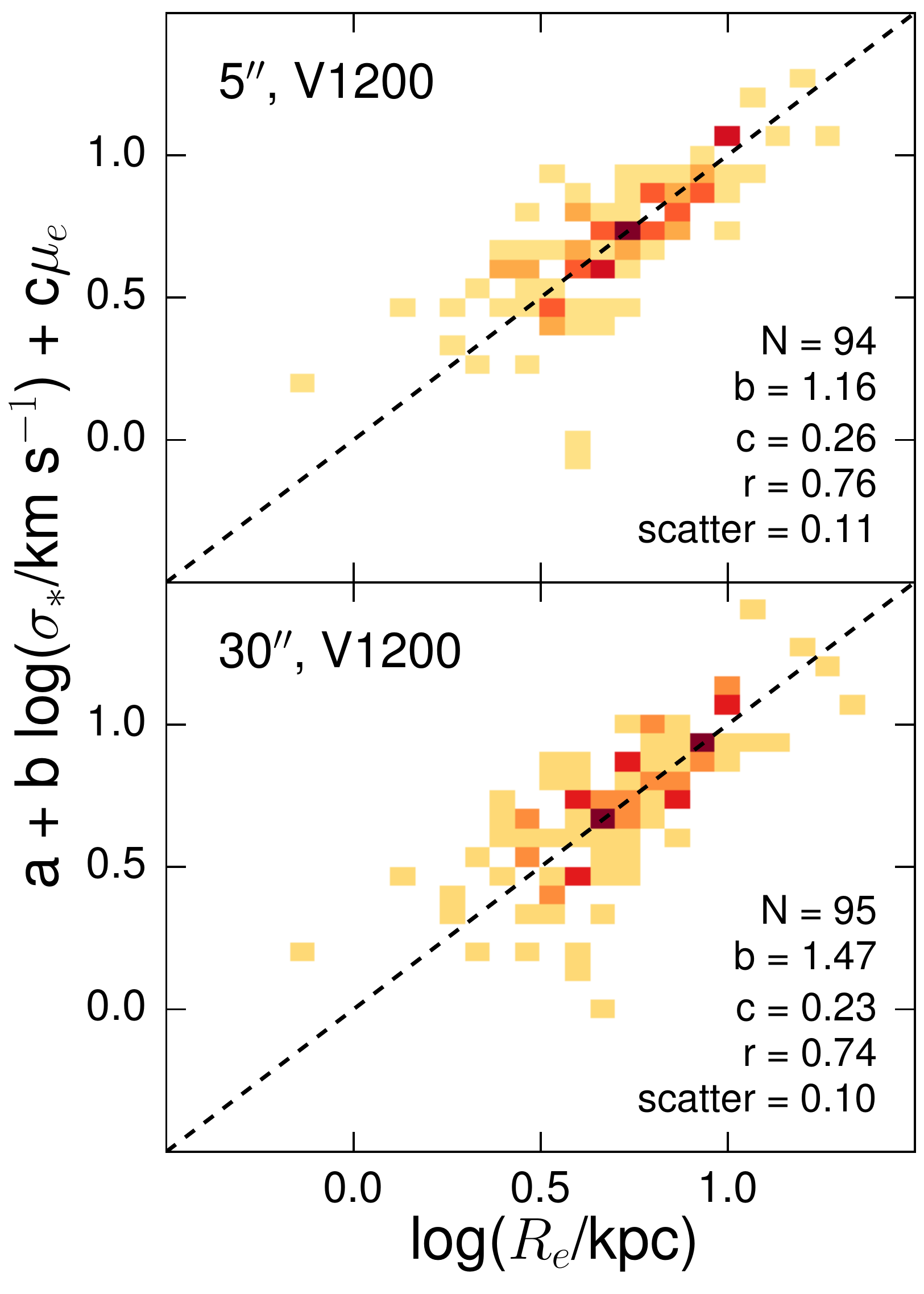}
\caption{The CALIFA Fundamental Plane, viewed edge-on, generated using Pipe3D V1200 stellar
velocity dispersions. $R_e$ and $\mu_e$ are sourced from the \protect\cite{gilhuly18} 
CALIFA photometry catalog. The dashed line corresponds to the fitted edge-on plane.}
\label{fig:FP}
\end{centering}
\end{figure}

We consider the Fundamental Plane (FP), the characteristic dynamical scaling relation of early-type
galaxies \citep{djorgovski87, dressler87, cappellari13, donofrio17}. We adhere to the completeness 
limits of CALIFA \citep[$-19 > M_r > -23.1$, see][]{walcher14}, and retain only elliptical and 
lenticular galaxies in the FP sample. \fig{FP} shows that the two FPs constructed using each of our 
stellar velocity dispersions successfully produce thin planes. 

The slope parameters for the FPs are broadly consistent with the literature. For instance, $b$ is 
typically measured as 1.06 - 1.55 \citep{dressler87, djorgovski87, bernardi03, labarbera08, cappellari13, ouellette17}, 
encompasing our values of 1.16 and 1.47. $b$ is typically found to decrease with larger apertures 
\citep{ouellette17}; however, our V1200 30$''$ aperture FP $b$ is closest to literature FPs 
constructed using the central velocity dispersion $\sigma_0$ \citep{bernardi03} while our V1200 
5$''$ aperture FP $b$ is similar to FPs constructed with $\sigma$ measured within $R_e$ 
\citep{cappellari13, ouellette17}. Considering the conversion from magnitudes to log $I_e$, typical 
values of $c$ are 0.30 - 0.36. Our values of 0.23 and 0.27 may lie outside this range due to bandpass 
differences or the properties of our ETG sample (especially the number of galaxies and the mass range 
investigated). Overall, this agreement is encouraging given the small sample size.

% Figure 7

%______________________________________________________________

\subsection{$S_{K}$ and the Tully-Fisher relation}
\label{sec:S05}

% Figure 8
\begin{figure}
\begin{centering}
\includegraphics[width=0.49\textwidth]{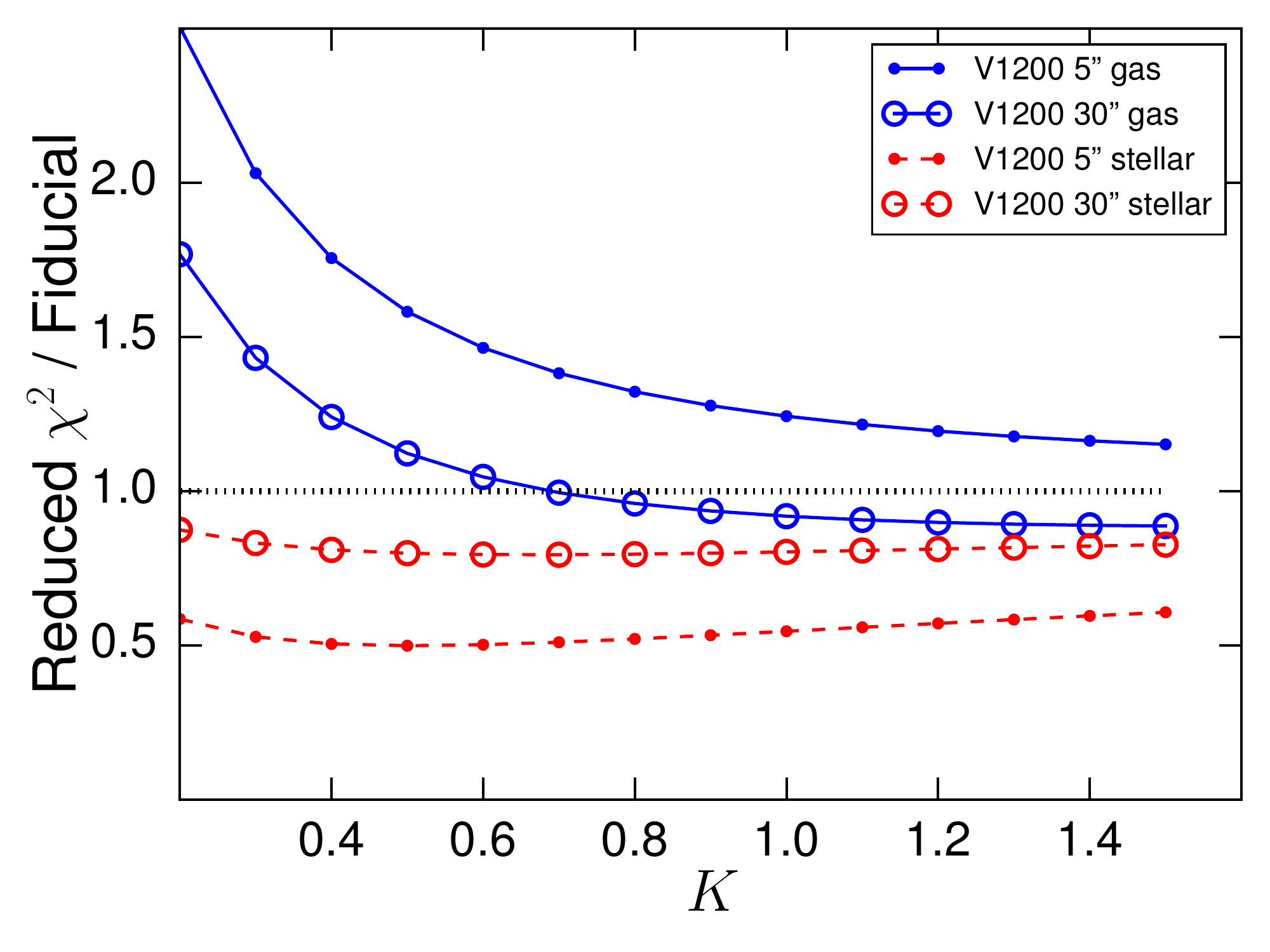}
\caption{Change in the reduced $\chi^2$ of the $M_* - S_K$ relation with increasing $K$ for
each dispersion, normalized against the reduced $\chi^2$ of the stellar mass TFR (i in dotted black).
Only the V1200 stellar dispersions (dashed red) provide significant improvement over 
the stellar mass TFR. In this work, $S_K$ is always constructed with $V_{max}$ measured from H$\alpha$ rotation curves.}
\label{fig:r_vs_Sk}
\end{centering}
\end{figure}

% Figure 9 (page-spanning)
\begin{figure*}
\begin{centering}
\includegraphics[width=0.9\textwidth]{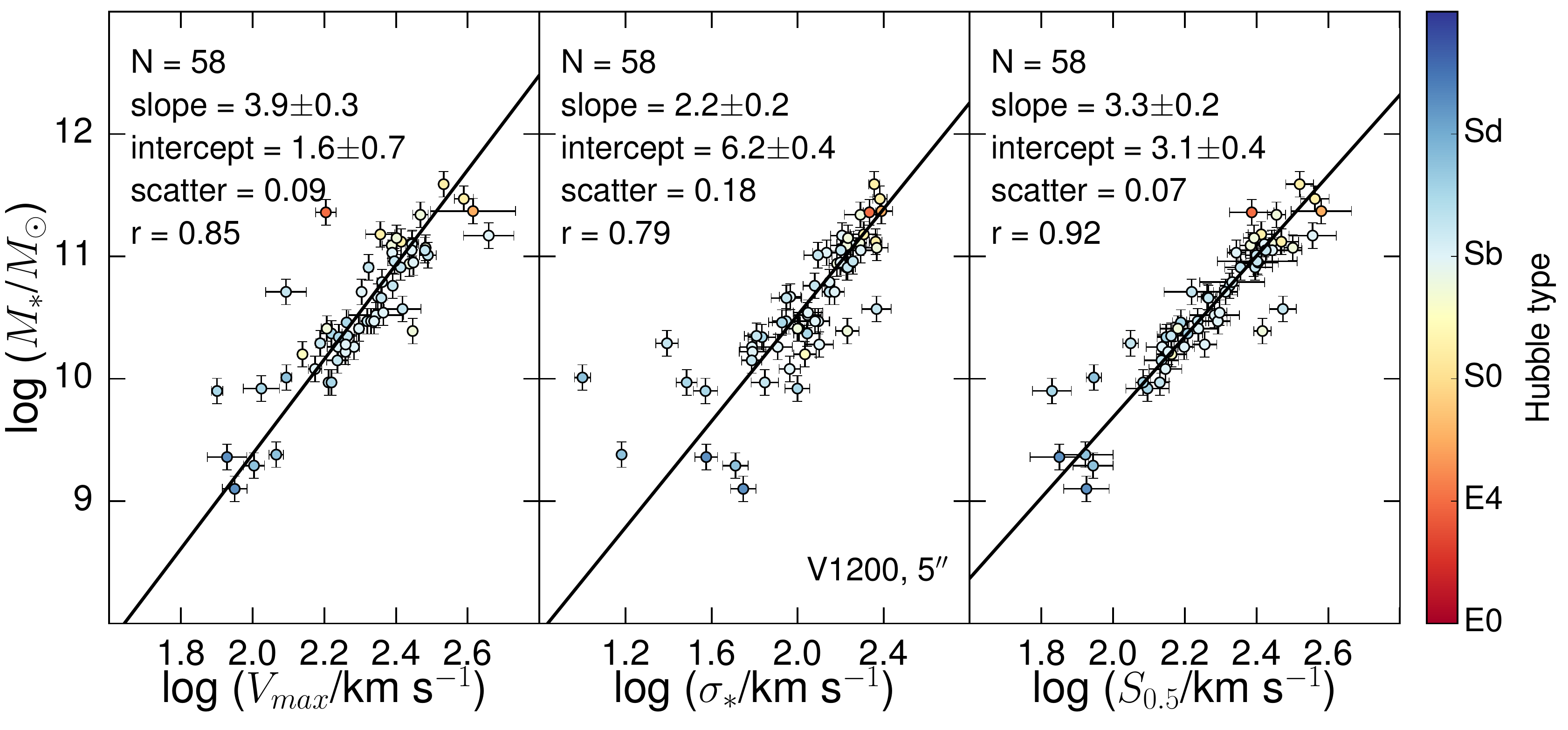}
\caption{A revised TFR using $S_{K}$ as the kinematic tracer (shown in the right panel). 
In the left panel, we show our stellar mass TFR constructed using $V_{max}$ \citep{holmes15}. 
The centre panel shows the $M_* - \sigma$ relation using the V1200 5$''$ stellar velocity 
dispersion. The final panel depicts the $M_* - S_K$ relation, taking $K=0.8$ (the value that 
results in the lowest reduced $\chi^2$). The solid lines show the best linear fits to the data. 
All points are colour-coded by Hubble type \citep{walcher14}.}
\label{fig:TF_S05_star}
\end{centering}
\end{figure*}

Unlike ETGs, LTGs are rotation-dominated and their respective dynamical scaling relations take 
advantage of circular velocities. Therefore, the ranking and exploitation of these dispersions 
becomes more complicated for LTGs. We now turn to the Tully-Fisher relation which connects the 
circular rotation velocity $V_{rot}$ and luminosity or stellar mass of disc galaxies 
\citep[TFR: ][]{tully77,mcgaugh00,courteau07}.

The potential of a dynamical system can be inferred by considering all stellar and gas motions.  
$S_K$ (Equation~1) could be used as a complement to pure circular velocity (typically for the gas) 
to account for random (stellar) motions and potentially yield a tighter scaling relation with $M*$
than the classic stellar mass TFR. While the most commonly-used value for $K$ is 0.5 
\citep{weiner06, kassin07, cortese14}, we test $0.2 \le K \le 1.5$ in increments of 0.1. The 
motivation for $K=0.5$ comes from the expected relation between velocity dispersion and rotation 
velocity for a population of tracers that have a spherically symmetric distribution and isotropic
velocity dispersion. If these conditions are met and $\rho \propto r^{-2}$, $K = \frac{1}{2}$.
One might also argue for $K=0.33$ from a consideration of kinetic energy. A rotating disc has 
$E_K = \frac{1}{2} M v_{rot}^2$ while a pressure-supported spherical galaxy has 
$E_K = \frac{3}{2} M \sigma_{v}^2$. If support from both rotation and pressure are
considered, the relative weighting would therefore be 3 to 1, yielding $K=\frac{1}{3}$.

In order to apply Equation~1, we use $V_{max}$ from the gas (\ha) rotation curve analysis of 
\cite{holmes15} for galaxies within the completeness limits of the CALIFA Data Release 1. Although 
Holmes et al. focused on mid-inclination galaxies, we apply our own inclination cut of 
$40^\circ < i < 80^\circ$ to ensure that only the most reliable rotation curves are retained. As 
these rotation curves trace the gas emission, the sample is dominated by gas-rich spiral galaxies 
and early-type galaxies are poorly represented. 

Examination of the reduced $\chi^2$ of our various $M_* - S_K$ relations in \fig{r_vs_Sk} reveals 
that our V1200 gaseous dispersions combined with a gaseous $V_{max}$ fail to produce tighter linear 
correlations than the classic TFR; in fact they degrade the TFR, such that the ideal value of $K$ 
is as large as possible to minimize the contribution of $\sigma$. This is at least partially 
attributed to the morphologically-tight, spiral-dominated sample for which \ha{} rotation curves 
were available.

Of the 5$''$ and 30$''$ aperture V1200 stellar dispersions, the smaller aperture yields the greatest 
improvement in correlation strength relative to the TFR when combined with a gaseous $V_{max}$ to 
form $S_K$. The reduced $\chi^2$ of the fitted $M_* - S_K$ for both apertures reaches its minimum
for $K=0.5$, though there is very little variation over the interval $0.3 \le K \le 0.8$. A comparison 
of the stellar mass TFR and our tightest $M_* - S_K$ relation is shown in \fig{TF_S05_star}. The
measured orthogonal scatter from the best-fitting $M_* - S_K$ linear relation is reduced by 0.02 dex
relative to that of the stellar mass TFR. It must be noted that the scatter of the stellar mass TFR 
was already quite low due to the spiral-dominated nature of this sample. A diverse sample, inclusive 
of ETGs and spanning a wider range of masses, could yield a relatively tighter $M_* - S_K$ relation
(Aquino et al., submitted). 

\cite{cortese14} find that the $M_* - S_{0.5}$ relation constructed with stellar or gaseous kinematics 
results in a scaling relation that is tighter than the baryonic TFR or the Faber-Jackson relation for 
235 SAMI galaxies of mixed morphology. They state that the scatter does not change for $0.3 < K < 1$. 
Our results also corroborate that a wide range of $K$ values yield similarly tight $M_* - S_K$
relations. Our measured slope ($3.3\pm 0.2$) and that of Cortese et~al. ($3.0 \pm 0.1$) are
consistent within error.

The galaxies studied by Cortese et~al. have lower stellar masses and velocities than ours; this 
difference is significant since the relative contribution of ordered and dispersive stellar motions  
changes for less massive galaxies. The differences between CALIFA and SAMI observations (spectral
resolution and coverage; spaxel size; radial coverage) and the techniques for measuring a rotation 
velocity may also contribute to our different results. For instance, we combine gas rotation 
velocities with stellar velocity dispersions while Cortese et~al. do not. Furthermore, our gaseous 
and stellar velocity dispersions clearly differ; for Cortese et al., $S_{0.5,gas}$ and $S_{0.5,*}$ 
are a close match (see their Fig.~3(c)). Similarly, we do not consider any stellar rotation
velocity for the construction of $S_K$, which penalises our analysis against early-type galaxies. 

Our tests have exploited the CALIFA library for well-behaved low redshift galaxies, and adopting
$S_{K}$ instead of $V_{rot}$ is likely less impactful than for high redshift and/or less massive galaxies.
Furthermore, we are limited in our ability to test $S_K$  for the unification of early- and late-type
galaxies by the small number of ETGs with significant ordered rotation traced in \ha{} emission
\citep{holmes15}. A larger sample with $V_{rot}$ measured for the stellar component in addition
to \ha{} may be needed to complete this study of $S_K$.

%______________________________________________________________

\section{Summary}
\label{sec:summary}

A table of dispersions measured in emission and absorption, tracing ionized gas and stars respectively,
has been presented for CALIFA galaxies. The data can be retrieved at this website:
\url{https://www.physics.queensu.ca/Astro/people/Stephane_Courteau/gilhuly2017/index.html}. 

Four different types of measurements are available, either from stellar absorption or gaseous
emission and using either 5$''$ or 30$''$ apertures. We note that every type of measurement may 
not be available for each galaxy. 

We have attempted to better contrast the distinguishing features of these dispersion measurements. 
Dispersions measured in emission or absorption trace different emitters and thus different dynamics. 
Galaxies with younger stellar populations (roughly $Z< 1$ and $A < 10^9$ yrs) may suffer from poorly 
constrained stellar dispersions and require special attention.

The Fundamental Planes constructed with our V1200 stellar dispersions agree well with the ensemble
of related studies in the literature, and no strong preference for the 5$''$ or 30$''$ aperture is
found for this purpose. We have also considered a modified TFR where $V_{rot}$ is replaced by the 
virial estimator $S_K$ for each of our dispersions and for $0.2 \le K \le 1.5$. Our measured 
dispersions do not result in a greatly tighter $M* - S_K$ relation than the classic TFR, but this 
may be related to the limited nature of our spiral-dominated sample, with its small mass range and 
lack of a stellar rotation velocity to compliment the \ha{} rotation curves of \cite{holmes15}. The 
$S_K$ produced with 5$''$ V1200 stellar velocity dispersions result in an $M_* - S_K$ relation with 
the most improved goodness of fit over the classic TFR (\fig{TF_S05_star}). Using these dispersions, 
we find that $0.3 \le K \le 0.8$ produce the tightest relations. The improvement in scatter would be 
better probed with a wider mass range and better representation of gas-poor systems.

%______________________________________________________________

\section*{Acknowledgments}

\noindent
We thank L. Holmes for providing us with machine-readable files containing the rotation curves 
presented in \cite{holmes15}. CG and SC acknowledge support from the Natural Science and 
Engineering Research Council (NSERC) of Canada through a PGS D scholarship and a Research 
Discovery Grant, respectively. SFS thanks the CONACyT program's CB-285080 and DGAPA-UNAM 
IA101217 grants for their support to this project.

%%%%%%%%%%%%%%%%%%%%%%%%%%%%%%%%%%%%%%%%%%%%%%%%%%


\bibliographystyle{mnras}
\bibliography{CALIFA_dispersions}

%%%%%%%%%%%%%%%%%%%%%%%%%%%%%%%%%%%%%%%%%%%%%%%%%%

%%%%%%%%%%%%%%%%% APPENDICES %%%%%%%%%%%%%%%%%%%%%

% N/A

%%%%%%%%%%%%%%%%%%%%%%%%%%%%%%%%%%%%%%%%%%%%%%%%%%

% Don't change these lines
\bsp	% typesetting comment
\label{lastpage}
\end{document}